\documentclass[aps, prd, twocolumn,showkeys,
showpacs,amsmath,amssymb,superscriptaddress]{revtex4-1}

\usepackage{graphicx}
\usepackage{dcolumn}
\usepackage{bm}

\newcommand{\dij}{\delta_{ij}}

\newcommand{\normaeqz}[1]{a^{(0 #1)}}
\newcommand{\normaeqo}[1]{a^{(1 #1)}_{i}}
\newcommand{\normaeqt}[1]{a^{(2 #1)}_{ij}}
\newcommand{\sign}{\sigma_{ij}} %weights as \omega
\newcommand{\aone}{\alpha^{(1)}}
\newcommand{\atwo}{\alpha^{(2)}}
\newcommand{\btwo}[1]{\beta^{(2|#1) }}

\newcommand{\bit}{\begin{itemize}}
\newcommand{\eit}{\end{itemize}}
\newcommand{\ben}{\begin{enumerate}}
\newcommand{\een}{\end{enumerate}}

\newcommand{\be}{\begin{equation}}
\newcommand{\ee}{\end{equation}}
\newcommand{\bee}{\begin{eqnarray}}
\newcommand{\eee}{\end{eqnarray}}

\newcommand{\ba}{\begin{align}} %ungleich aligned!!!, array gibts auch noch
\newcommand{\ea}{\end{align}}
\newcommand{\bc}{\begin{center}}                 
\newcommand{\ec}{\end{center}}
\newcommand{\bn}{\\\\$\begin{aligned}\centering}
\newcommand{\en}{\end{aligned}$}
\newcommand{\bal}{\begin{aligned}}
\newcommand{\eal}{\end{aligned}}

\newcommand{\tx}[1]{\text{#1}}

\newcommand{\ff}[2]{\frac{#1}{#2}}

\newcommand{\vv}[1]{\mathbf{#1}} %\newcommand{\vv}[1]{\boldsymbol{#1}}

\newcommand{\exv}[1]{\langle #1 \rangle}

\newcommand{\on}[1]{\operatorname{#1}}

\newcommand{\q}{\mathbf{q}}

\newcommand{\td}{\tx{d}}
\newcommand{\te}{\tx{e}}

\newcommand{\g}{\gamma}

\newcommand{\de}{\delta}

\begin{document}

\title{Gaussian Quadrature and Lattice Discretization of the \\Fermi-Dirac
  Distribution for Graphene}

\author{D. Oettinger} 
\affiliation{ETH Z\"urich, Department of
  Physics, CH-8093 Z\"urich, Switzerland } 

\author{M. Mendoza} \email{mmendoza@ethz.ch} \affiliation{ ETH
  Z\"urich, Computational Physics for Engineering Materials, Institute
  for Building Materials, Schafmattstrasse 6, HIF, CH-8093 Z\"urich
  (Switzerland)}

\author{H. J. Herrmann} \affiliation{ ETH Z\"urich, Computational
  Physics for Engineering Materials, Institute for Building Materials,
  Schafmattstrasse 6, HIF, CH-8093 Z\"urich (Switzerland)}
\affiliation{Departamento de F\'isica, Universidade Federal do
  Cear\'a, Campus do Pici, 60455-760 Fortaleza, Cear\'a, (Brazil)}

\date{\today}

\begin{abstract}
  We construct a lattice kinetic scheme to study electronic flow in
  graphene. For this purpose, we first derive a basis of orthogonal
  polynomials, using as weight function the ultrarelativistic
  Fermi-Dirac distribution at rest. Later, we use these polynomials to
  expand the respective distribution in a moving frame, for both
  cases, undoped and doped graphene. In order to discretize the
  Boltzmann equation and make feasible the numerical implementation,
  we reduce the number of discrete points in momentum space to $18$ by
  applying a Gaussian quadrature, finding that the family of
  representative wave (2+1)-vectors, that satisfies the quadrature,
  reconstructs a honeycomb lattice. The procedure and discrete model
  are validated by solving the Riemann problem, finding excellent
  agreement with other numerical models. In addition, we have extended
  the Riemann problem to the case of different dopings, finding that
  by increasing the chemical potential, the electronic fluid behaves
  as if it increases its effective viscosity.
\end{abstract}

\maketitle

\section{Introduction}

Since its discovery \cite{natletter,Geim1}, graphene has shown a
series of wonderful electrical and mechanical properties, such as
ultra-high electrical conductivity, ultra-low viscosity, as well as
exceptional structural strength, combined with mechanical flexibility
and optical transparence. Due to the special symmetries of the
honeycomb lattice, electrons in graphene are shown to behave like {\it
  massless} chiral ultrarelativistic quasiparticles, propagating at a
Fermi speed of about $v_F \sim 10^6$ m/s \cite{DiVincenzo.1984,
  revDasSarma}. This places graphene as an appropriate laboratory for
experiments involving relativistic massless particles confined to a
two-dimensional space \cite{PhysToday}.

Electronic gas in graphene can be approached from a hydrodynamic
perspective \cite{Muller.2009, grapPRB, grap2PRB, grap3PRB}, behaving
as a nearly perfect fluid reaching viscosities significantly smaller
than those of superfluid Helium at the lambda-point. This has
suggested the possibility of observing pre-turbulent regimes, as
explicitly pointed out in Ref. \cite{grapPRL} and later confirmed by
numerical simulations \cite{turbPRL}. All these characteristics in
graphene open up the possibility of studying several phenomena known
from classical fluid dynamics, e.g. transport through disordered media
\cite{SR2013}, Kelvin-Helmholtz and Rayleigh B\'enard instabilities,
just to name a few. However, the study of these phenomena needs
appropriate numerical tools, which take into account both, the
relativistic effects and the Fermi-Dirac statistics.

Recently, a solver for relativistic fluid dynamics based on a minimal
form of the relativistic Boltzmann equation, whose dynamics takes
place in a fully discrete phase-space lattice and time, known as
relativistic lattice Boltzmann (RLB), has been proposed by Mendoza et
al.~\cite{rlbPRL, rlbPRD} (and subsequently revised in
Ref.~\cite{rlbhupp} enhancing numerical stability). This model
reproduces correctly shock waves in quark-gluon plasmas, showing
excellent agreement with the solution of the full Boltzmann equation
obtained by Bouras et al. using BAMPS (Boltzmann Approach Multi-Parton
Scattering) \cite{BAMPS, BAMPSs}. In order to set up a theoretical
background for the lattice version of the relativistic Boltzmann
equation for the Boltzmann statistics, Romatschke et
al. \cite{PRCpaul} developed a scheme for an ultrarelativistic gas
based on the expansion in orthogonal polynomials of the
Maxwell-J\"uttner distribution \cite{RelaBoltEqua} and, by following a
Gauss-type quadrature procedure, the discrete version of the
distribution and the weight functions was calculated. This procedure
was similar to the one used for the non-relativistic lattice Boltzmann
model \cite{HERMI, HERMI2, He.1997, Succi.2001}. This relativistic
model showed very good agreement with theoretical data, although it
was not compatible with a lattice, thereby requiring linear
interpolation in the free-streaming step. Another model based on a
quadrature procedure was developed recently in order to make the
relativistic lattice Boltzmann model compatible with a lattice
\cite{rlb_diss}. However, all these models are based on the the
Maxwell-J\"uttner distribution, which is based on the Boltzmann
statistics, and therefore, their applications to quantum systems is
limited.

In this work, we construct a family of orthogonal polynomials by using
the Gram-Schmidt procedure using as weight function the
ultrarelativistic Fermi-Dirac distribution at rest. By applying a
Gauss-type quadrature, we find that the family of discrete
(2+1)-momentum vectors, needed to recover the first three moments of
the equilibrium distribution, are fully compatible with a hexagonal
lattice, avoiding any type of linear interpolation. This result is
very convenient, since the crystal of graphene shares the same
geometry, facilitating the implementation of boundary conditions,
allowing for instance having a good approximation for the electronic
transport in nanoribbons with armchair or zigzag edges
\cite{GNR1,GNR2} by implementing the typical bounce-back rule for
lattice Boltzmann models.

The paper is organized as follows: in Sec. \ref{methods}, we describe
in details the expansion of the Fermi-Dirac distribution in an
orthogonal basis of polynomials, and perform the Gauss-type
quadrature. In this section, we also explain the discretization
procedure. In Sec.~\ref{sec:num}, we implement the validation of our
model by simulating the Riemann problem; and in Sec.~\ref{sec:num1},
we perform additional simulations for doped graphene. Finally, in
Sec.~\ref{discussions}, we discuss the results and future work.

\section{Model Description}\label{methods}

The electronic gas in graphene can be considered as a gas of massless
Dirac quasi-particles obeying the Fermi-Dirac statistics in a
two-dimensional space. Thus, we define the single-particle
distribution function $f(x^{\mu}, p^{\mu})$ in phase space, being
$x^{\mu}=(x^0, x^1, x^2)$ and $p^\mu = (p^0, p^1, p^2)$ the
time-position and energy-momentum coordinates, respectively. Here
$x^0$ denotes time, $\vec{x} = (x^1, x^2)$ spatial coordinates, $p^0$
the energy, and $\vec{p}=(p^1, p^2)$ the momentum of the particles. In
the ultrarelativistic regime, we get $p^\mu p_\mu = 0$ (in this paper
we use the Einstein notation, i.e. repeated indexes denote summing
over such indexes). In our approach, we assume that the distribution
function $f$ evolves according to the relativistic Boltzmann-BGK
equation \cite{RelaBoltEqua},
\begin{equation} 
  p^{\mu}\partial_{\mu} f = -\ff{p_\alpha U^\alpha}{v_F^2 \tau}(f-f_{eq}) \quad,
\end{equation}
where $\tau$ is the relaxation time, and $f_{eq}$ the equilibrium
distribution, which in our case, is the relativistic Fermi-Dirac
distribution defined by
\begin{equation}\label{fermimi} 
  f_{eq}(x^\mu,p^\mu)=\ff{1}{\te^{(p_{\alpha}U^{\alpha}-\mu)/k_B T} + 1} \quad ,
\end{equation}
with $T$ the temperature, $k_B$ the Boltzmann constant, $U^{\mu}$ the
macroscopic (2+1)-velocity of the fluid \cite{Juttner.1928,
  RelaBoltEqua}, and $\mu$ the chemical potential. The relation
between the Lorentz-invariant $U^{\mu}$ and the classical velocity
$\vec{u}=(u^1,u^2)$ is given by $U^{\mu}=\g(v_F,u^1,u^2)$, with $v_F$
being the Fermi speed and $\gamma=1/\sqrt{1-\vec{u}^2/v_F^2}$.
 
\subsection{Moment expansion}

Here, we perform an expansion of the Fermi-Dirac distribution,
Eq.~\eqref{fermimi}, in an orthogonal basis of polynomials. In our
case, since we are interested in the hydrodynamic regime, we will
truncate the expansion preserving only the polynomials up to second
order, although achieving higher orders is also possible by using the
same procedure. In particular, we need to reproduce the first three
moments of the equilibrium Fermi-Dirac distribution, namely
$\exv{1}_{(eq)}$, $\exv{p^{\alpha}}_{(eq)}$, and
$\exv{p^{\alpha}p^{\beta}}_{(eq)}$ for $\alpha,\beta=0,1,2$. The
angular brackets denote expectation values using the distribution $f$
via $\exv{Q}=\int \td\mu\, Q f$, with $\td\mu=d^2 p /2p^0(2\pi)^2$,
and the subscript $_{(eq)}$ indicates that the equilibrium
distribution $f_{eq}$ is taken instead of $f$.
  
This method was originally introduced by Grad \cite{Grad.1949} who
expanded the Maxwell-Boltzmann distribution in Hermite polynomials,
based on the fact that they are orthogonal, using as weight function
the Maxwellian distribution at rest. In this spirit, we will derive a
new basis of polynomials that are orthogonal with respect to the
Fermi-Dirac distribution at rest,
\begin{equation}
  w(p_0)=\ff{1}{\te^{p_0/k_B T} + 1} \quad .
\end{equation}
  
For the following derivations it is useful to choose natural units,
$c=k_B=\hbar=1$. In addition, we will consider only the case for $\mu
= 0$, although a general approach is straightforward. By introducing a
reference temperature $T_0$, we define $\theta = T/T_0$, $\bar p =
p^0/T_0$, $\vec{v} = \vec{p}/|\vec{p}|$, and using $p^0 = |\vec{p}|$,
we rewrite the equilibrium distribution as
 \begin{equation}\label{feq} 
   f_{eq,E}(t,\vec{x},\bar p, \vec{v})=\ff{1}{\te^{\bar p \gamma ( 1-
       \vec{v} \cdot \vec{u})/\theta} + 1 } \quad ,
\end{equation}
where the subscript $E$ stands for ``Exact''. The distribution
$f_{eq,E}$ is expanded using tensorial polynomials $P^{(n)}$, for the
angular contribution, and $F^{(k)}$, for the radial dependence, such
that
\begin{equation}\label{ansatz}
  f_{eq,E}(t,\vec{x},p, \vec{v})=\ff{1}{\te^{\bar
      p}+1}\sum_{n,k}^{\infty}a^{(nk)}_{\underline i}(t,\vec{x})P^{(n)}_{\underline
    i}(\vec{v})F^{(k)}(\bar p) \quad .
\end{equation} 
Here, the (2+1)-momentum vectors have been expressed in polar
coordinates, $p^\mu = (\bar p, \bar p \cos\phi, \bar p \sin\phi )$
with $\vec{v} = (\cos\phi, \sin\phi)$ being a unit vector that carries
the angular dependence $\phi$, and the index $\underline i$ denotes a
family of indices $i_1,...,i_n\in\{1,2\}$ whose total number equals
the order $n$ of the tensor for the angular dependence,
i.e. $P^{(n)}_{\underline i}$ and $a^{(nk)}_{\underline i}$ are
tensors of rank $n$. Such an ansatz has been used by Romatschke et
al. \cite{PRCpaul} to expand the Maxwell-J\"uttner
distribution. Employing the Gram-Schmidt procedure, the radial
polynomials $F^{(k)}$ are constructed satisfying the orthogonality
relation
\begin{equation} 
  \int_0^{\infty} \ff{\td \bar p}{4\pi} w(\bar p) F^{(k)}(\bar
  p)F^{(l)}(\bar p) = \Gamma^{(k)}_{F}\de_{kl} \quad ,
\end{equation}
and the angular ones by satisfying
\begin{equation}
  \int_0^{2\pi}\ff{\td\phi}{2\pi}P^{m}_{\underline
    i}(\phi)P^{n}_{\underline j}(\phi)=\Gamma^{(m)}_{P,\underline i
    \underline j}\de_{mn} \quad .
\end{equation}
The resulting polynomials and $\Gamma$-constants up to second order
are given in Appendix ~\ref{appendixA}.  With these polynomials and
taking into account Eq.~\eqref{ansatz}, one can show that up to second
order in $n$ and $k$, we get
\begin{equation}\label{eq:ank} 
  a^{(nk)}_{\underline i}=\ff{g^{(n)}}{\Gamma_F^{(k)}}T_0\int \ff{\td \bar
    p}{4\pi}\ff{\td\phi}{2\pi} f_{eq,E} P^{(n)}_{\underline
    i}F^{(k)} \quad , 
\end{equation} 
with $g^{(0)}=1$, $g^{(1)}=2$, and $g^{(2)}=4$. The explicit form of
$a^{(nk)}$ is given in Appendix~\ref{appendixB}. Using
Eq.~\eqref{ansatz}, the definitions of the polynomials, and their
orthogonality relations it can be easily shown that the moments up to
second order can be written in terms of the coefficients
$a^{(nk)}_{\underline i}$ with $n, k \leq 2$ (see
Appendix~\ref{appendixB}), and therefore, the truncated expansion of
the distribution $f_{eq}$ up to second order becomes
\begin{equation}\label{eq:truncated}
  f_{eq}=\ff{1}{\te^{\bar p}+1}\sum_{n=0}^{2}\sum_{k=0}^{2} a^{(nk)}_{\underline
    i}P^{(n)}_{\underline i}F^{(k)} \quad .
\end{equation}
This is sufficient to recover the moments %%
\begin{subequations}
  \begin{equation}
    \exv{p^{\alpha}}_{(eq)}=n U^{\alpha} \quad , \label{eq:moments1}
  \end{equation}
\begin{equation}
  \exv{p^{\alpha}p^{\beta}}_{(eq)}= (\epsilon + P) U^{\alpha}U^{\beta} -
  P \eta^{\alpha\beta } \quad , \label{eq:moments2}
\end{equation} 
\end{subequations}
of the full Fermi-Dirac distribution, Eq.~\eqref{feq}. In
Eq.~\eqref{eq:moments2}, we have introduced the Minkowski metric
tensor $\eta^{\alpha\beta }$, the particle density
$n=\ff{\pi}{48}T^2$, the pressure $P=\ff{9\zeta(3)}{\pi^2}nT$ and the
energy density $\epsilon=2P$, where $\zeta$ denotes the Riemann zeta
function, $\zeta(3)\approx1.202$.
\begin{figure}
\centering
\includegraphics[width=0.515\linewidth]{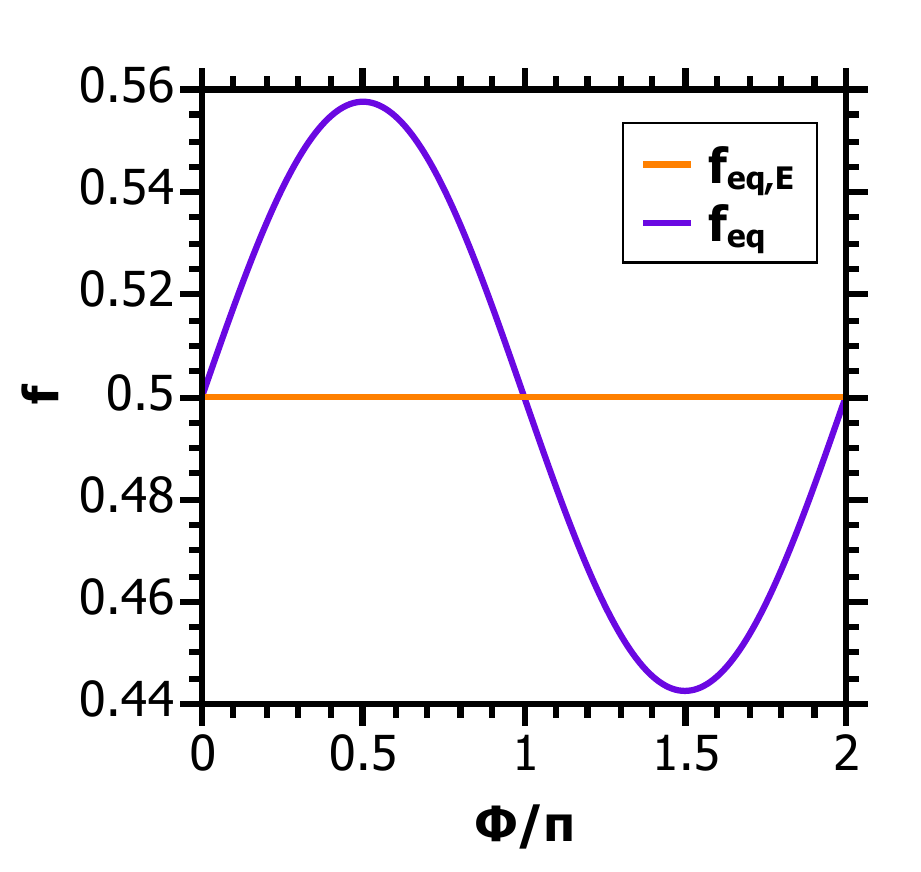}\includegraphics[width=0.515\linewidth]{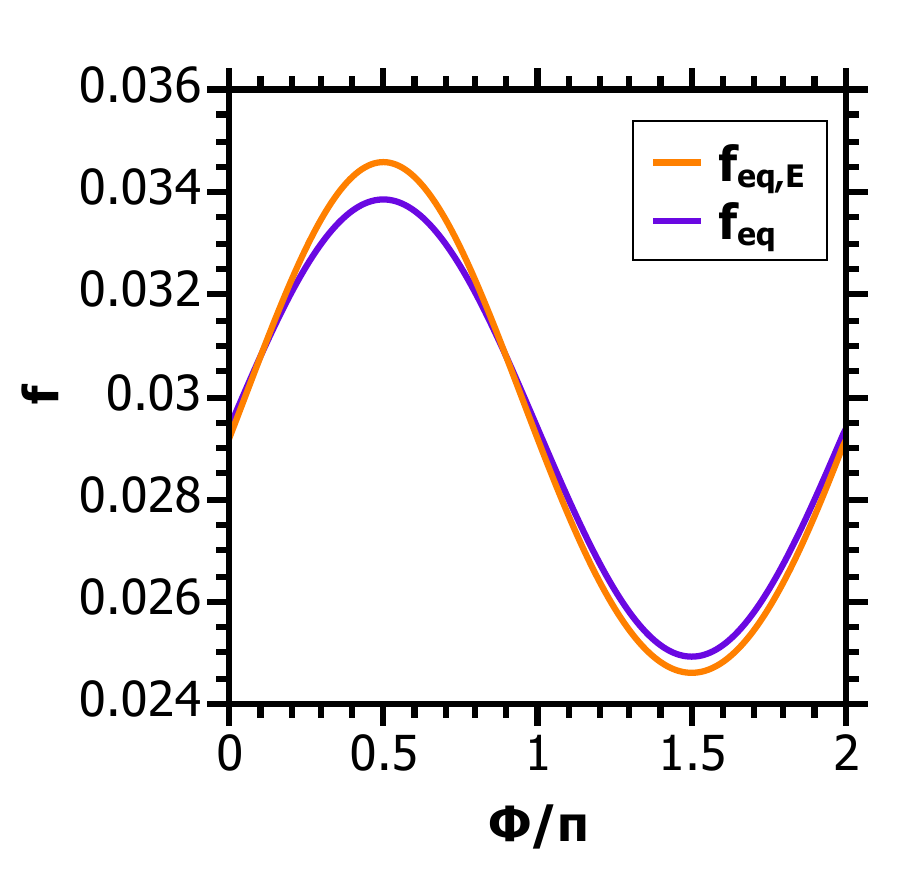}
\includegraphics[width=0.515\linewidth]{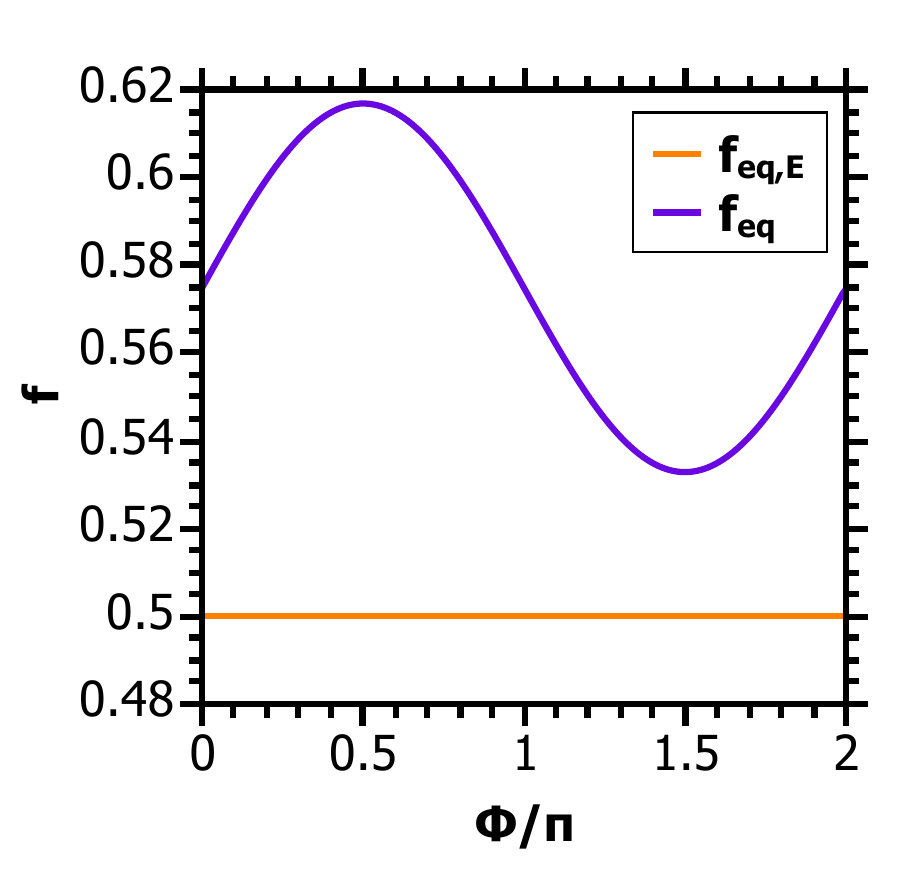}\includegraphics[width=0.515\linewidth]{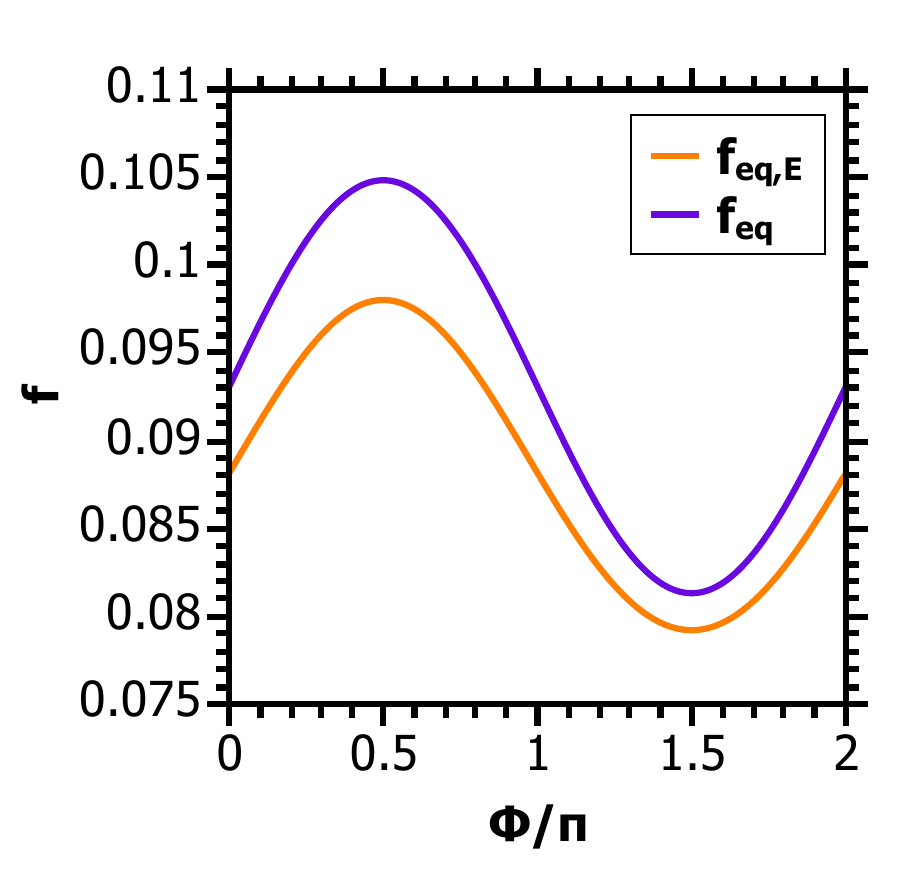}
\caption{\label{fig:matching02} Comparison betweem the expanded Fermi
  dirac distribution $f_{eq}$ and the full version $f_{eq, E}$ as a
  function of the angular component $\phi$, for $\bar p=0$ (left) and
  $\bar p=3.5$ (right), with $\theta=1.0$ (top) and $\theta=1.5$
  (bottom), $u^1=0.0$, $u^2=0.05$.}
\end{figure}

Fig.~\ref{fig:matching02} shows that the quality of the matching
between the truncated $f_{eq}$ and the exact $f_{eq,E}$, for $\bar p
\sim 0$, is very poor, in contrast with the case, $\bar p \sim
3.5$. However, this is not surprising, since we are dealing with a gas
of ultrarelativistic particles which are always moving at the Fermi
speed, and therefore none of them has energy $\bar p = 0$. On the
other hand, the matching is reasonable for $\theta=1$, while being off
for $\theta > 1$. Thus, we conclude that $\theta=1$ offers the best
approximation, and therefore, we will work with that value. In
addition, we have found that $\theta$ cannot be chosen far below unity
because $f_{eq}$ can present negative values. The fact that $\theta =
1$ implies that the reference temperature $T_0$ should be equal to the
temperature of the electronic gas $T$.

\subsection{Momentum space discretization}

We now need to discretize the momentum space into a finite number $N$
of discrete momentum vectors, $p^{\mu}_q$ (with $q = 0, ..., N$) such
that we can replace integrals in the continuum momentum space by sums
over a small number of discrete momentum (2+1)-vectors. In order to do
that, we use the Gaussian quadrature \cite{Davis.1984}. As an example,
for the radial dependence of the expansion, in order to satisfy
\begin{equation} 
  \int_0^{\infty} \ff{\td\bar p}{4\pi} w(\bar p) F^{(k)}(\bar p) \bar p^l = \sum_{q'=0}^N
  \ff{\omega^{(\bar p)}_{q'}}{w(\bar p_{q'})}w(\bar p_{q'})
  F^{(k)}(\bar p_{q'}) \bar p^l_{q'} \quad ,
\label{eq:radialquadrature} 
\end{equation}
for $k,l\leq 2$, we should calculate the discrete $\bar p_{q'}$ and
respective radial weights $\omega^{(\bar p)}_{q'}$. By using the
Gaussian quadrature theorem, we found the following values:
\begin{eqnarray*}
  \bar p_1=0.484,&&~\omega^{(\bar p)}_{1}= 0.0369\\
  \bar p_2= 2.447,&&~\omega^{(\bar p)}_{2}= 0.0176\\
  \bar p_3= 6.424,&&~\omega^{(\bar p)}_{3}= 0.000719 \quad .
\end{eqnarray*}
Note that in fact, $\bar p$ is always larger than zero, as expected
for ultrarelativistic particles, (see Appendix \ref{appendixC} for
numerical values with higher precision).

On the other hand, by following a similar procedure, we can calculate
the $N'$ discrete angles $\phi_{q''}$ and angular weights
$\omega^{(\phi)}_{q''}$ (with $q'' = 1,..., N'$), such that, for the
angular integrals over $P^{(n)}(v_i)^l(v_j)^m$, one gets
\begin{equation}
  \int_{0}^{2\pi}\ff{\td\phi}{2\pi}P^{(n)}(v_i)^l(v_j)^m = \sum_{q''=0}^{N'}
  \omega^{(\phi)}_{q''} P^{(n)}(v_{i, q''})^l(v_{j, q''})^m \quad ,
  \label{eq:angularquadrature}
\end{equation}
where $v_{i, q''}$ denotes $v_i(q'')$. The above expression is
required to be an exact quadrature formula for $n\leq 2$, and
$l+m\leq2$. The results for the discrete angles and weights functions
are $\phi_{q''}=\ff{\pi}{2}+(q''-1)\ff{\pi}{3}$ and
$\omega^{(\phi)}_{q''}=\ff 16$ with $N'=6$.  

\begin{figure}
  \includegraphics[width=0.8\linewidth]{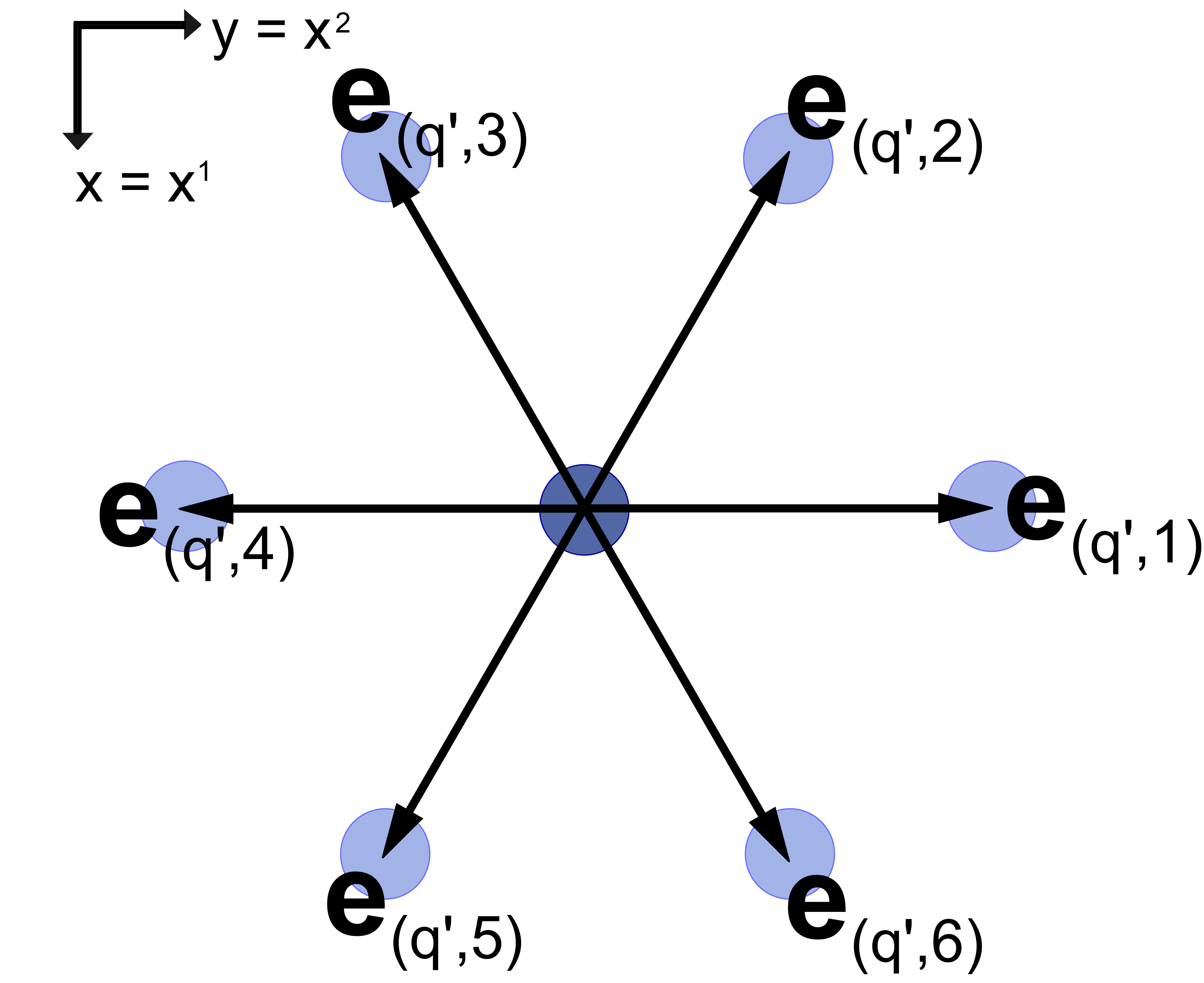}
  \caption{\label{fig:lb_hexagonal_lattice} The populations $f_{\vv
      q}$ are moved between the nodes of a hexagonal lattice which are
    linked by the vector $\vec{e}_{\vv q}\de t$.}
\end{figure}

By combining the radial and angular dependence of the discrete
momentum (2+1)-vectors we get a total of $18$ discrete lattice vectors
$p^{\mu}_{\vv q}=p^{\mu}_{(q',q'')}=T_0(\bar p_{q'}, \bar p_{q'}
\cos\phi_{q''},\bar p_{q'} \sin\phi_{q''} )$, where we have introduced
the index $\vv q=(q',q'')$. This lattice cell configuration is shown
in Fig.~\ref{fig:lb_hexagonal_lattice}, where we can observe that for
recovering hydrodynamics in graphene, we need a hexagonal
lattice. This is a very convenient result, since due to the fact that
it possesses the same honeycomb lattice symmetries of graphene, we can
reproduce with good accuracy boundary conditions when modeling
nanoribbons or other complex structures.

The exact quadrature relations, Eqs.~\eqref{eq:radialquadrature} and
\eqref{eq:angularquadrature}, ensure that the moments up to second
order are still represented exactly: 
\begin{subequations}
\begin{equation}
  \exv{p^{\alpha}}_{(eq)}=\sum_{\vv q} \ff{\omega_{\vv q}}{w(\bar
    p_{\vv q})} f_{(eq),\vv q} p^{\alpha}_{\vv
    q} \quad , 
\label{eq:moments1discrete}
\end{equation}
\begin{equation}
  \exv{p^{\alpha}p^{\beta}}_{(eq)}=\sum_{\vv q} \ff{\omega_{\vv
      q}}{w(\bar p_{\vv q})} f_{(eq),\vv q} p^{\alpha}_{\vv
    q}p^{\beta}_{\vv q} \quad . 
\label{eq:moments2discrete} 
\end{equation}
\end{subequations}

We have expanded and discretized the Fermi-Dirac equilibrium
distribution for ultrarelativistic particles. Now, we will proceed to
discretize the Boltzmann equation and find the evolution equation for
the non-equilibrium distribution.

\subsection{Lattice Boltzmann algorithm}

With the expanded distribution functions and the discretization of
momentum space at hand, we may use the following discrete Boltzmann
equation \cite{He.1997,PRCpaul, rlbPRD},
\begin{equation} 
  f_{\vv q}(t+\de t, \vec{x} + \vec{e}_{\vv q}\de t) - f_{\vv q}(t, \vec{x}) =
  -\ff{p^\alpha U_\alpha}{p^0\tau}(f_{\vv q}(t, \vec{x}) - f_{eq,\vv q}(t, \vec{x})) , 
\label{eq:LBequation}
\end{equation} 
where we have introduced the notations $\vec{e}_{\vv q}=\vec{p}_{\vv
  q}/p^0$, and $f_{\vv q}(t,\vec{x})= f(t,\vec{x}, p_{\vv q})$. Note
that $\vec{e}_{\vv q}$ are unit vectors, which means that there are
effectively $6$ different $\vec{e}_{\vv q}$. The discrete Boltzmann
equation is now embedded into a lattice, and each time step of $\de t=
1$ corresponds to one execution of the following steps:
\begin{enumerate}
\item Calculate the equilibrium distributions $f_{eq,\vv
    q}(t,\vec{x})$ from Eq.~\eqref{eq:truncated} using the macroscopic
  variables $n=n(t,\vec{x})$, $\vec{u}=\vec{u}(t,\vec{x})$, and
  $T(t,\vec{x})$. At $t=0$, $n(t=0, \vec{x})$, $T(t=0, \vec{x})$, and
  $\vec{u}(t=0,\vec{x})$ are imposed as initial conditions.
\item Collision: Introducing the post-collisional distributions
  $f'_{\vv q}$, calculate \[ f'_{\vv q}(t,\vec{x})=f_{\vv
    q}(t,\vec{x})-\ff{p^\alpha U_\alpha}{p^0\tau}(f_{\vv q}(t,
  \vec{x})-f_{eq, \vv q}(t, \vec{x}) ). \] At $t=0$, take $f_{\vv
    q}=f_{eq, \vv q}$.
\item Streaming: Move the $f'_{\vv q}$ along $\vec{e}_{\vv q}$:
  \[ f_{\vv q}(t+1,\vec{x} + \vec{e}_{\vv q}) =f'_{\vv q}(t,
  \vec{x}) \]
\item Calculate the new macroscopic variables. First we compute the
  energy density of the system by solving the eigenvalue problem,
  $\langle p^\alpha p^\beta \rangle U_\alpha = \epsilon U^\beta$,
  according to the Landau-Lifshitz decomposition
  \cite{RelaBoltEqua}. From this, we get $\epsilon$ and
  $U^\alpha$. Next, we use the relation $n = \langle p^\alpha \rangle
  U_\alpha = n$ to obtain the particle density. Here, the average
  values, $\langle p^\alpha \rangle$ and $\langle p^\alpha p^\beta
  \rangle$, are simply
  \begin{eqnarray*} \langle p^\alpha \rangle &=&\sum_{\vv q}
    \ff{\omega_{\vv q}}{w(\bar p_{\vv q})} f_{\vv q} p^\alpha_{\vv q}
    \quad , \\
    \langle p^\alpha p^\beta \rangle &=& \sum_{\vv q} \ff{\omega_{\vv
        q}}{w(\bar p_{\vv q})} f_{\vv q} p^\alpha_{\vv q} p^\beta_{\vv
      q} \quad .
  \end{eqnarray*}
\end{enumerate}
The streaming step indicates that if we discretize the real space
based on a hexagonal lattice where the sites are linked by
$\vec{e}_{\vv q}\de t$, as shown in
Fig.~\ref{fig:lb_hexagonal_lattice}, the values of $f_{\q}$ will be
moved between these sites exactly. This is known as ``exact
streaming'' and crucial for the computational efficiency and accuracy
of the lattice Boltzmann methods, because it removes any spurious
numerical diffusivity.

In summary, we have developed a (2+1)-dimensional relativistic lattice
Boltzmann scheme with the remarkable feature that it takes into
account the Fermi-Dirac statistics, while recovering all the moments
up to second order. The discretization is realized on a hexagonal
lattice such that exact streaming is achieved. The fact that the
quadrature corresponds to a hexagonal lattice allows to represent
complex boundaries more precisely in graphene applications. This will
be studied in more details in future works.

Up to now, we are working with undoped graphene, $\mu = 0$. However,
by using the same orthogonal polynomials, we can easily integrate the
Fermi-Dirac statistics for the doped case, obtaining the extended
formulation. In this work, we will use $\mu = 0$, in order to compare
the results with previous models in the literature that use the
Maxwell-J\"uttner distribution, since transport theory shows that in
the case of undoped Fermi-Dirac statistics, the transport
coefficients, namely shear viscosity and thermal conductivity, have
the same expresions than for the Boltzmann statistics
\cite{RelaBoltEqua}. Therefore, the shear viscosity takes the value of
$\eta = (3/5) P (\tau - \delta t/2)$ \cite{jstat2013}. Later, we will
use the doped case to study the Riemann problem, which to best of our
knowledge has never been studied before. However, it is present when,
for instance, laser beams are pointed to the graphene sheet in order
to measure transport coefficients \cite{laserbeams}.

\section{Validation: Riemann problem}\label{sec:num} 

\begin{figure}
\includegraphics[width=1.07\linewidth]{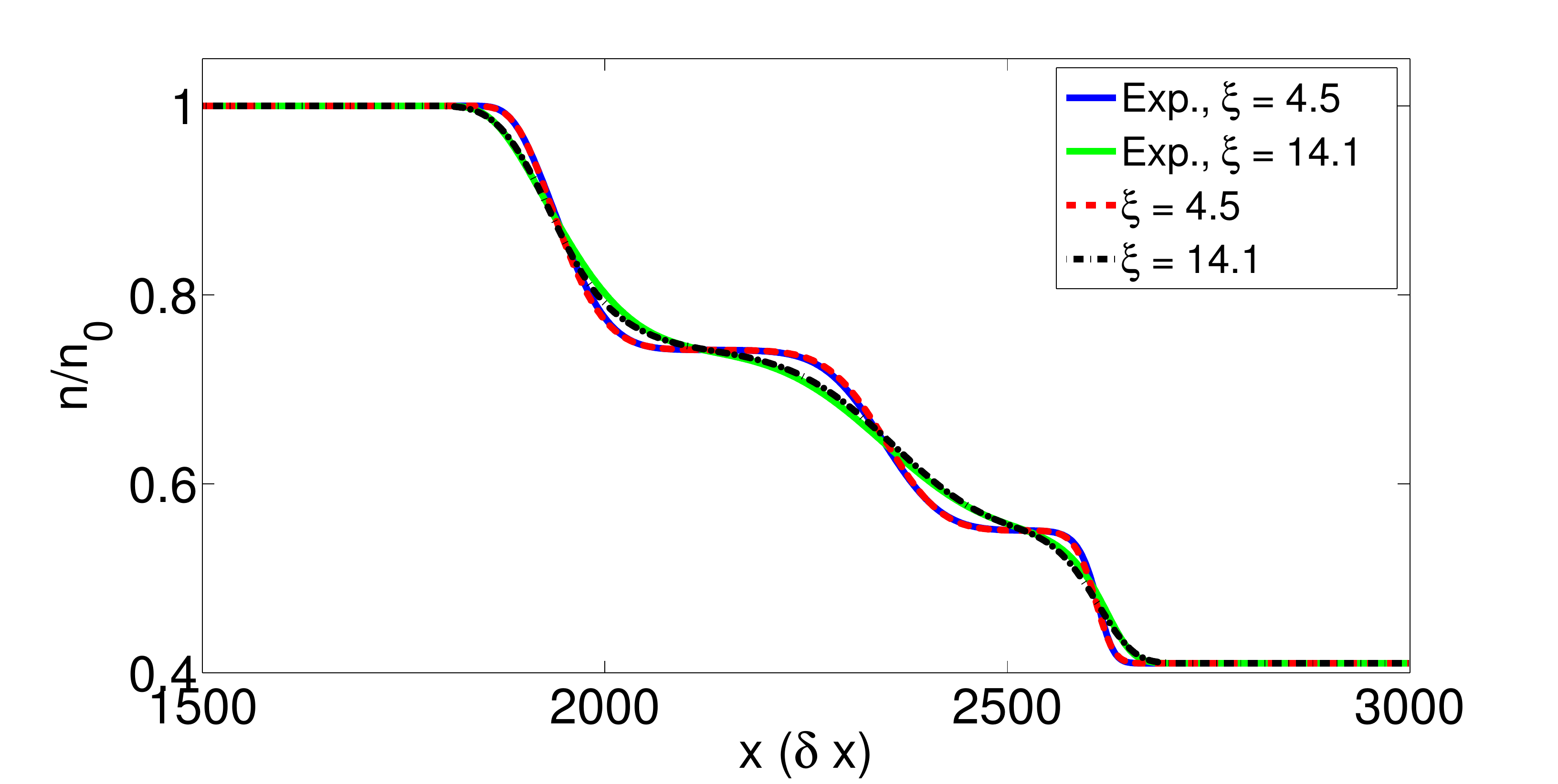}
\includegraphics[width=1.07\linewidth]{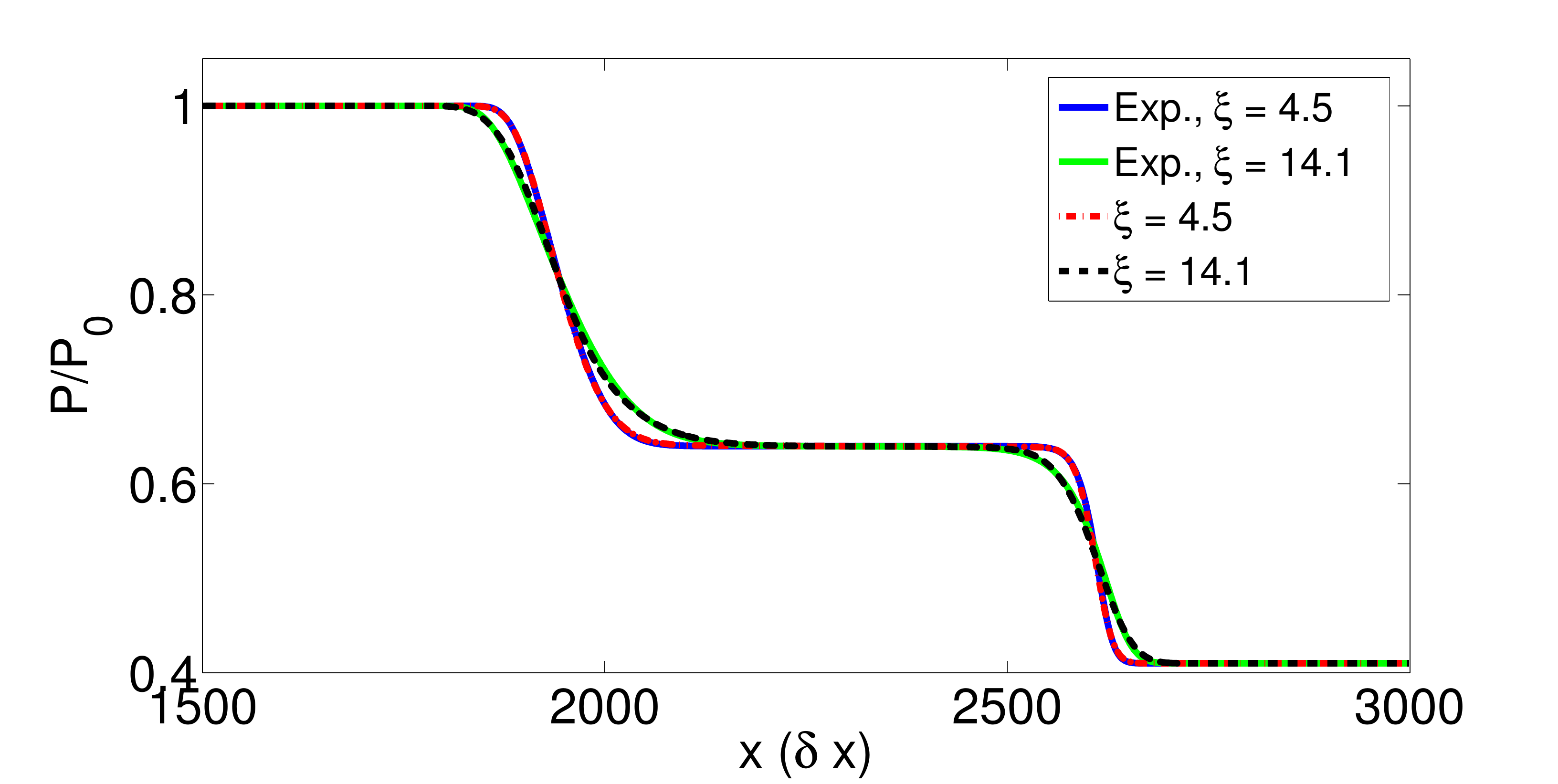}
\includegraphics[width=1.07\linewidth]{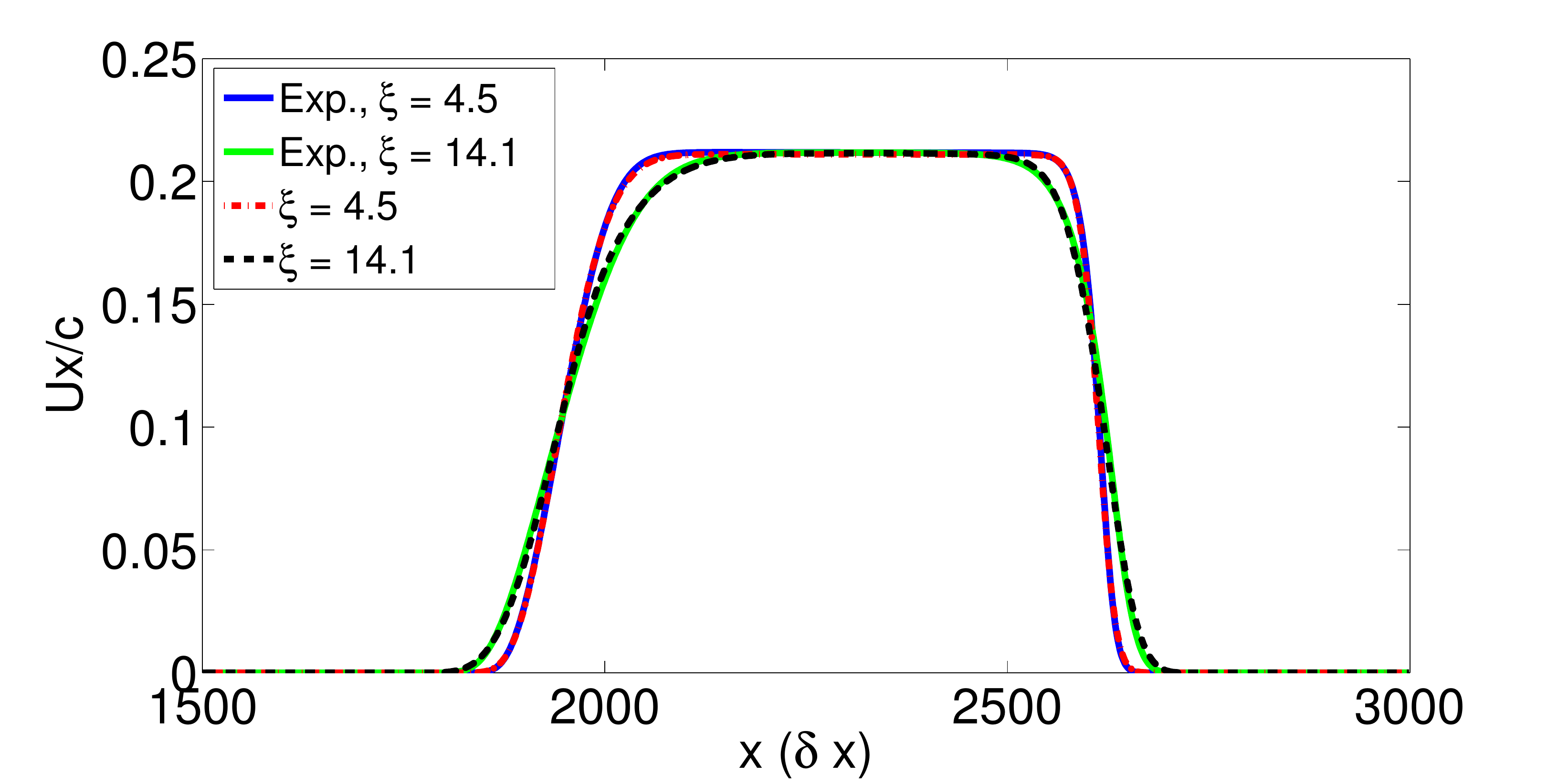}
\caption{\label{fig:density_riemann} Density, pressure and velocity
  profile for the solution of the Riemann problem. Here $\xi =
  \eta/(P_0 \delta t)$ is a dimensionless number. The expected results
  were calculated using the model in Ref.~\cite{rlb_diss}.}
\end{figure}

In order to validate our model, we solve the Riemann problem for the
ultrarelativistic Fermi-Dirac gas. The Riemann problem is a standard
test for both, relativistic and non-relativistic hydrodynamics
numerical schemes, because it involves the evolution of two states of
the fluid initially separated by a discontinuity. In our case, we set
up an effectively one-dimensional system of $L_x\times L_y =
3000\times 2$ nodes, using periodic boundary conditions in $x$ and $y$
components. Initially, there are two regions with particle densities,
$n_0 = 1$ ($3 L_x/4 > x > L_x/4$), and $n_1 = 0.41$ ($x \le L_x/4$ and
$x \ge 3L_x/4$) creating a rectangular plateau of non-zero particle
density in the center of the simulation zone. Here we consider and
initial constant temperature, $T_0 = 1$. The initial velocity is set
to zero and the value of the relaxation time $\tau$ is calculated for
two different values of $\xi = \eta/(P_0 \delta t)$, with
$P_0=\ff{9\zeta(3)}{\pi^2}n_0 T_0$. The evolution of the system is
displayed in Fig.~\ref{fig:density_riemann} after $470$ time steps,
showing the generated shock wave. We have only plotted the region
$x>L_x/2$ since the other one does not give additional
information. Note that there is excellent agreement with the solutions
provided by the model proposed in Ref.~\cite{rlb_diss} for the same
initial conditions.

\section{Riemann problem with $\mu \neq 0$}\label{sec:num1} 

Let us now consider the case when the chemical potential $\mu$ is
different from zero. For this purpose, we follow the same procedure
described before but this time, we keep $\mu \neq 0$. The development
is straightforward, and therefore does not deserve a full
explanation. The polynomials are the same as described in Appendix
\ref{appendixA}, and the coefficients $a^{(nk)}_{\underline i}$ are
calculated by using Eq.~\eqref{eq:ank}. 

The hydrodynamic approach of electrons in graphene works for low
doping, $\mu/k_B T \ll 1$ \cite{grapPRB, grap2PRB,
  grap3PRB}. Therefore, we can expand the discrete equilibrium
distribution in powers of $\mu/k_BT$ up to third order, neglecting
errors of the order of $(\mu/k_B T)^4$. We perform additional
simulations of the Riemann problem with the same parameters as before,
but now, varying the chemical potential.
\begin{figure}
\includegraphics[width=1.06\linewidth]{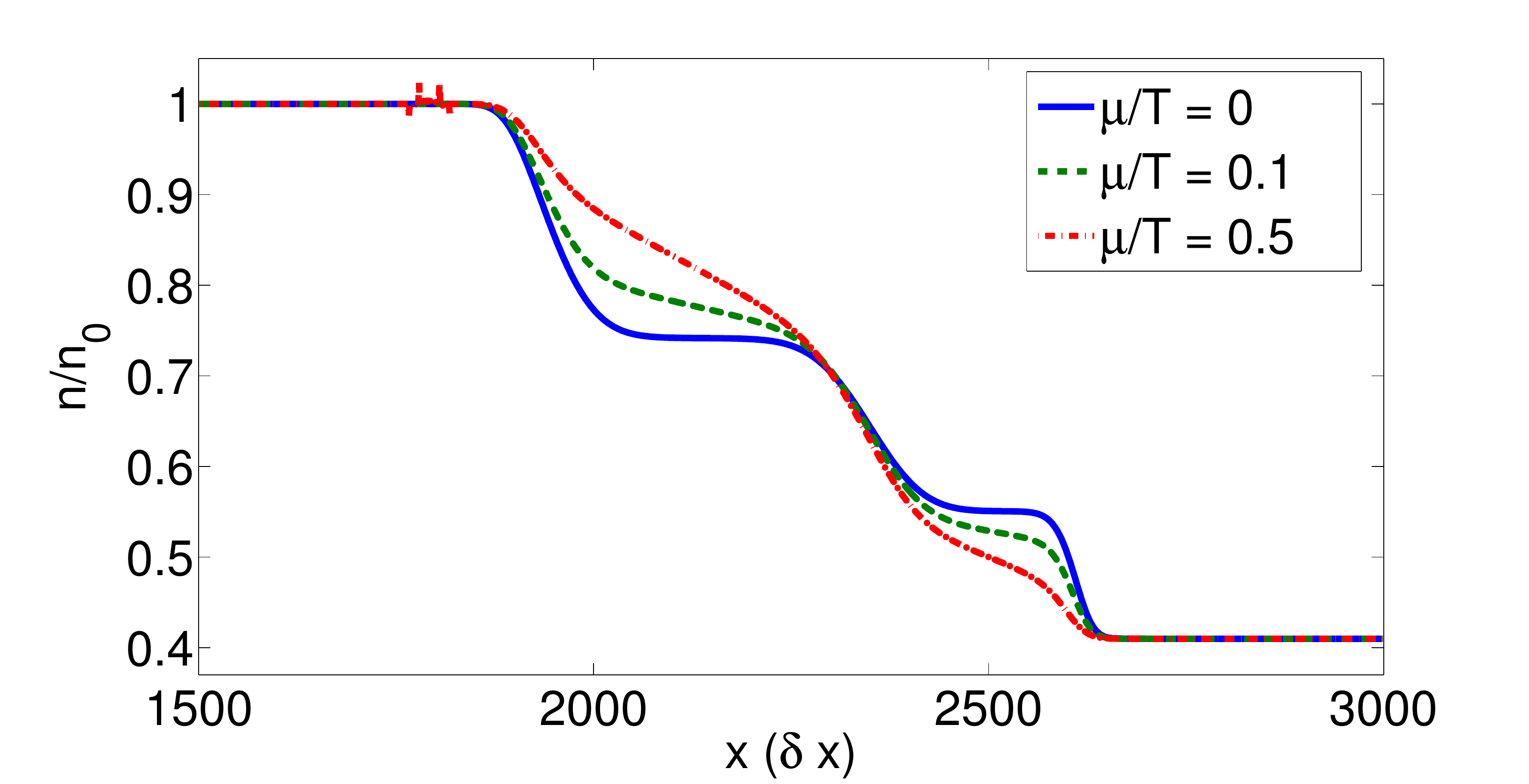}
\includegraphics[width=1.06\linewidth]{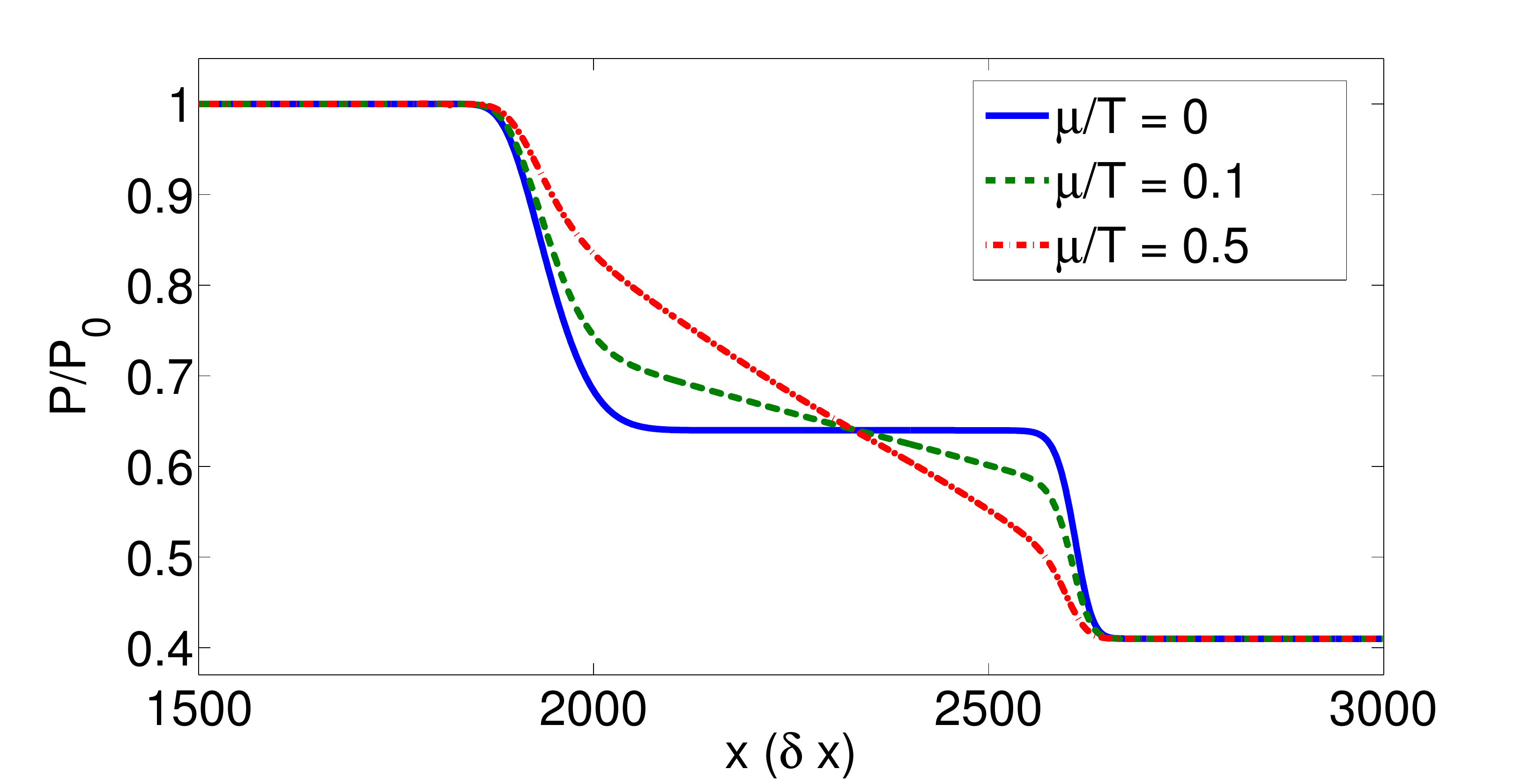}
\includegraphics[width=1.06\linewidth]{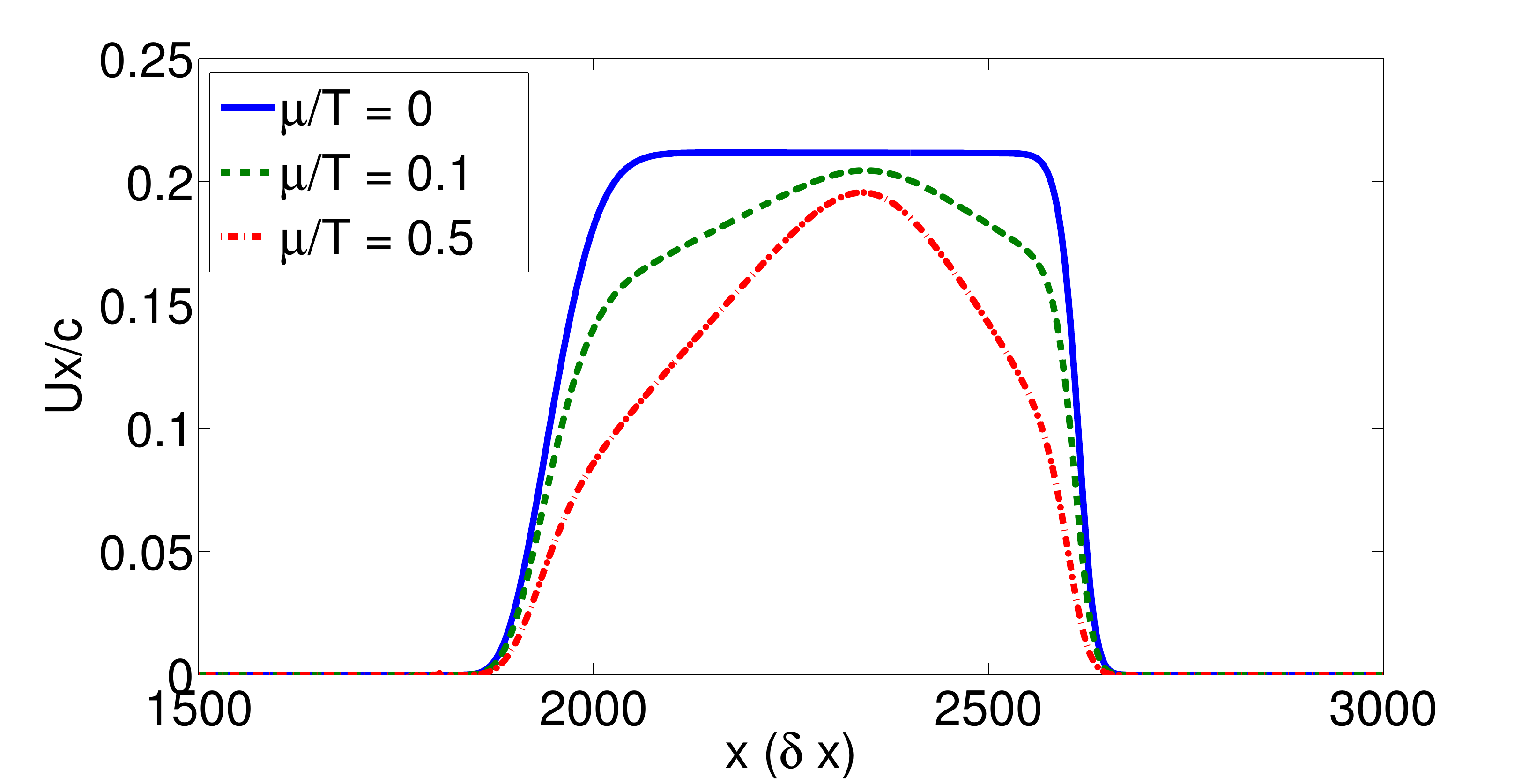}
\caption{\label{fig:density_riemann2} Density, pressure and velocity
  profile of the solution of the Riemann problem, for different values
  of the chemical potential $\mu$.}
\end{figure}
As we can observe from Fig.~\ref{fig:density_riemann2}, increasing the
chemical potential tends to increase also the effective viscosity of
the system, smoothing the profiles of the velocity, pressure and
density. This result is very interesting because it suggest that, in
fact, impurities with soft potentials (small $\mu/k_B T$) in graphene
samples can be treated as local modifications in the effective
viscosity of the electronic fluid. In other words, this result
suggests a promissing way to include impurities in the hydrodynamic
approach of electrons in graphene. Note that in this figure, there is
a noise in the profile of the particle density. This numerical
instability remains with the same amplitude and is always located at
the boundary when $n=n_0$, and therefore, it does not destroy the
stability of the simulation. It can be due to the relevance of higher
order terms which are not recovered by our expansion.

\section{Conclusions}\label{discussions}

We have derived a new family of orthogonal polynomials using as weight
function the Fermi-Dirac distribution for ultrarelativistic particles
in two dimensions. By applying the Gaussian quadrature we have
calculated the set of representative momentum (2+1)-vectors, which
allows us to replace the integrals over the continuum momentum space
by sums over such vectors. As a very interesting result, we have found
that those vectors possess the same symmetries than the honeycomb
lattice of carbon atoms in graphene, making possible the accurate
implementation of complex boundary conditions in future applications,
such as point defects and nanoribbons. The derivation has been
performed by imposing that the expanded distribution should fulfill at
least the first three moments of the equilibrium distribution, which
are needed to recover the appropriate hydrodynamics. However, higher
order moments can also be recovered by using the same procedure in
this paper.

In addition, we have developed a new lattice kinetic scheme to study
the dynamics of the electronic flow in graphene. The model is
validated on the Riemann problem, which is one of the most challenging
tests in numerical hydrodynamics, presenting excellent agreement with
previous models in the literature. By increasing the chemical
potential, we have found that the profiles of velocity, particle
density, and pressure, change similar to the case when the viscosity
is increased, concluding that increasing the Fermi energy results in
increasing the effective viscosity of the electronic fluid.  This
result suggests that soft impurities in graphene samples can be
treated as local modifications of the viscosity, however, further
studies must be performed in order to confirm this statement.

The fact that we can propagate the information from one site to
another in an exact way, avoiding interpolation, removes any kind of
spurious numerical diffusivity. Therefore, we expect this model to be
appropriated to study many problems in electronic transport in
graphene in the framework of the hydrodynamic approach,
e.g. turbulence and hydrodynamical instabilities in graphene flow,
just to name a few.

Extensions of the present model to take into account higher order
moments of the Fermi-Dirac equilibrium distribution as well as the
inclusion of the distribution and dynamics of holes, will be a subject
of future research.

\begin{acknowledgements}
  We acknowledge financial support from the European Research Council
  (ERC) Advanced Grant 319968-FlowCCS.
\end{acknowledgements}

\appendix
\section{Polynomials and $\Gamma$-constants}\label{appendixA}

In this section, we write explicitly the family of polynomials, which
are orthogonal using as weighting function the Fermi-Dirac
distribution at rest, with their respective normalization factors. For
the case of the angular dependence, we have
\begin{eqnarray*}
P^{(0)}(\vec{v}) &=& 1 \\
P^{(1)}_i(\vec{v}) &=& v_i \\
P^{(2)}_{ij} (\vec{v}) &=& v_i v_j - \ff 12 \dij 
\end{eqnarray*}
with normalization factors,
\begin{eqnarray*}
\Gamma_P^{(0)}&=&1\\
\Gamma_{P,ij}^{(1)}&=&\ff 12 \dij \\
\Gamma_{P,ijkl}^{(2)}&=& \ff 18
\left(\delta_{il}\delta_{jk}+\delta_{ik}\delta_{jl}-\delta_{ij}\delta_{kl}\right)
\quad .
\end{eqnarray*}

For the case of the radial dependence, we have the polynomials
\begin{eqnarray*}
  F^{(0)}(\bar p)&=&1\\
  F^{(1)}(\bar p)&=&\bar p - c_{10} \quad , \\  c_{10}&=&\ff{\pi^2}{12\log(2)}
  \quad ,\\
  F^{(2)}(\bar p)&=&\bar{p}^2 -c_{21} \bar p - c_{20}\quad , \\
  c_{21}&=&- \ff{6 (7 \pi^4 \log(2) - 15 \pi^2 \zeta(3))}{5 (\pi^4 -
    216 \log(2) \zeta(3))}\quad , \\ c_{20}&=&\ff{7 \pi^6 - 3240
    \zeta(3)^2}{10 (\pi^4 - 216 \log(2) \zeta(3))} \quad ,
\end{eqnarray*}
with $\zeta$ denoting the Riemann zeta function. The normalization
factors for these polynomials are:
\begin{eqnarray*}
  \Gamma_F^{(0)}&=&\ff{\log(2)}{4\pi} \quad , \quad
  \Gamma_F^{(1)}=-\ff{\pi^3}{576 \log(2)} + \ff{3 \zeta(3)}{8\pi}
  \quad , \\
  \Gamma_F^{(2)}&=&\ff{1}{400 \pi}\biggl( \ff{49 \pi ^8 \log(2) - 210 \pi^6 \zeta(3) + 
    48600 \zeta(3)^3)}{\pi^4 - 216 \log(2)\, \zeta(3)} \\ &+& 2250 \zeta(5)
  \biggr ) \quad .
\end{eqnarray*}

\section{Coefficients for the expansion of $f_{eq}$ and relation to
  moments}\label{appendixB}

The coefficients of the expansion in Eq.~\eqref{eq:truncated} are
given by
\begin{eqnarray*}
  \normaeqz 0 &=& \theta \quad , \quad \normaeqo 0 = 2\,\theta\ff{1}{\g+1}\,u_i\g \quad ,\\
  \normaeqt 0 &=& \sign\,4\,\theta\frac{1}{(\g+1)^2}\left[\gamma^2(u_i
    u_j - \tfrac 12 \delta_{ij})\g^2 + \tfrac 12 \delta_{ij} \right]
  \quad , \\
  \normaeqz 1 &=& \aone \theta\left[\theta\g-1\right] \quad ,\\
  \normaeqo 1 &=& \ff{2\aone \theta}{\g+1}\left[\theta(\g+1)-1\right]\,u_i\g \quad ,\\
  \normaeqt 1 &=&\frac{4\aone \theta}{(\g+1)^2} \left[
    (2-\dij)\theta(\g+2)-\sign \right] [\gamma^4(u_i u_j -
  \delta_{ij}/2 ) \\ &+& \delta_{ij}/2 ] \quad ,\\
  \normaeqz 2 &=& \atwo \theta\, \big[\btwo 1 (\theta\g-1)+ \btwo 2
  ((3\theta\g-2)-\theta) \\ &+& \btwo 3 (\theta^2 (3\g^2-1)-2) \big]
  \quad , \\
  \normaeqo 2 &=& \ff{2 \atwo \theta}{\g+1}\big[ \btwo
  1(\theta(\g+1)-1) \\ &+& \btwo 2 \theta (3\theta\g-2)(\g+1) \\ &+&
  \btwo 3 (3\theta^2\g(\g+1)-2)  \big] u_i\,\g  \quad , \\
  \normaeqt 2 &=& \ff{4 \atwo \theta}{(\g+1)^2}\big[ \btwo 1( (2-\dij)\theta(\g+2)-(2\dij+1)\sign ) \\
  &+& \btwo 2 (3\theta^2(\g+1)^2-2(2-\dij)(\theta(\g+2)-2\dij\sign))  \\ 
  &+& \btwo 3 (3\theta^2(\g+1)^2-2\sign) \big] [\gamma^2(u_i u_j
  - \tfrac 12 \delta_{ij})\g^2 \\ &+& \tfrac 12 \delta_{ij} ] \quad ,  
  \nonumber\end{eqnarray*}
where $\sign=(-1)^{\de_{2,i}\de_{2,j}}$ or
\begin{equation}
  (\sign) =  \left( \begin{array}{cc}
      1 & 1 \\
      1 & -1 \end{array}\right) \quad ,
\end{equation}
and,
\begin{eqnarray*}
  \aone &=& \ff{12 \pi^2 \log (2)}{216\log(2)\zeta(3)-\pi^4} \quad ,\\
  \atwo &=& 5\,[2250\zeta(5)(216\log(2)\zeta(3)-\pi^4)+210\pi^6\zeta(3)
  \\ &-&49\pi^8\log(2) -48600\zeta(3)^3]^{-1} \quad ,\\
  \btwo 1 &=& -14\pi^6 \log(2) \quad ,\\
  \btwo 2 &=& -15\pi^4\zeta(3)^2 \quad ,\\
  \btwo 3 &=& 3240\log(2)\zeta(3)^2 \quad , \\
\end{eqnarray*}
which are approximately, $\aone \approx 0.994$, $\atwo \btwo 1 \approx
-1.629$, $\atwo \btwo 2 \approx -0.307$, and $\atwo \btwo 1 \approx
0.567$.

To obtain the moments from the expansion of $f_{eq}$, we expressed
them in terms of the $a^{(nk)}_{\underline i}$ using
Eqs.~\eqref{eq:ank}, \eqref{eq:truncated}, and the expressions in
Appendix \ref{appendixA}, e.g.
$\exv{p^0}=T_0^2\left(\Gamma_F^{(1)}\normaeqz 1 + c_{10}\Gamma_F^{(0)}
  \normaeqz 0 \right)$.
 
Note that for the calculation of the coefficients
$a^{(nk)}_{\underline i}$ we should use of the integration formula
\[ \int_0^{\infty}\td x \ff{x^{n-1}}{z^{-1}\te^{a\,x}+1} =
-z^{-1}a^{-n}\Gamma(n)\on{Li}_n(-z) \quad , \] which holds for $n>0$,
$a\in\mathbb R$, $a>0$.  Here, $\Gamma(n)$ denotes the gamma function,
which becomes $\Gamma(n)=(n-1)!$ for $n\in \mathbb N$. $\on{Li}_n(z)$
is the polylogarithm which can be defined using a power series:
$\on{Li}_n(z)=\sum_{k=1}^{\infty}\ff{z^k}{k^n}$.  If we consider the
chemical potential in the Fermi-Dirac distribution to be zero, we have
$z=1$ and the relevant values of the polylogarithm become
$\on{Li}_1(-1)=-\log(2),\,\on{Li}_2(-1)=-\ff{\pi^2}{12},\,\on{Li}_3(-1)=-\ff{3}{4}\zeta(3)$. On
the other hand, for $\mu \neq 0$, we take $z = e^{\mu/T}$.
 
\section{Results for radial Gaussian quadrature}\label{appendixC}

When the radial Gaussian quadrature is applied, the following values
for the discrete $\bar p_i$ are obtained:
 \begin{eqnarray*}
   \bar p_1&=& 0.4840534751554060637550794361591 \quad , \\
   \bar p_2&=& 2.4467448689670852668751189804200 \quad , \\
   \bar p_3&=& 6.4243522612255152565859012563254 \quad ,
 \end{eqnarray*}
with its respective weight functions
 \begin{eqnarray*}
   \omega^{(\bar p)}_{1}&=& 0.0368730611359638360101542425978  \quad , \\
   \omega^{(\bar p)}_{2}&=& 0.0175666801777458993453757617390 \quad , \\
   \omega^{(\bar p)}_{3}&=& 0.0007191587244531629935841036927 \quad .
 \end{eqnarray*}
\vspace{1cm}

\bibliography{LBgraphene_paper}

\end{document}